\pdfoutput=1

\documentclass[11pt]{article}

\usepackage[final]{acl}

\usepackage{times}
\usepackage{latexsym}

\usepackage[T1]{fontenc}

\usepackage[utf8]{inputenc}

\usepackage{microtype}

\usepackage{inconsolata}

\usepackage{graphicx}

\usepackage{makecell}
\usepackage{booktabs}
\usepackage{multirow}
\usepackage{xspace}
\usepackage{amsmath}
\usepackage{algorithm}
\usepackage{algpseudocode}
\usepackage{bbding}
\usepackage{arydshln}
\usepackage{bm}
\usepackage{xcolor}

\usepackage{ulem}

\newcommand{\model}{{CLMTracing}\xspace}

\usepackage{float}

\title{\model: Black-box User-level Watermarking \\ for Code Language Model Tracing}

\author{
    ~~Boyu Zhang$^{1}$\footnotemark[1]
   ~~Ping He$^{1}$\footnotemark[1]
   ~~Tianyu Du$^{1}$\footnotemark[2]
   ~~Xuhong Zhang$^{1}$
   ~~Lei Yun$^{2}$
   \AND
   ~~Kingsum Chow$^{1}$\footnotemark[2]
   ~~Jianwei Yin$^{1}$\\ 
   $^{1}$Zhejiang University 
   $^{2}$Information Security Center, CEPREI   \\
   \texttt{{\{zjuzby, gnip, zjradty, zhangxuhong\}@zju.edu.cn}}\\
   \texttt{{yl@ceprei.com, kingsum.chow@zju.edu.cn}}\\
\texttt{{zjuyjw@cs.zju.edu.cn}}\\
}

\begin{document}
\maketitle

\renewcommand{\thefootnote}{\fnsymbol{footnote}}
\footnotetext[1]{Equal contribution.}
\footnotetext[2]{Corresponding author.}
\renewcommand{\thefootnote}{\arabic{footnote}}

\begin{abstract}
With the widespread adoption of open-source code language models (code LMs), intellectual property (IP) protection has become an increasingly critical concern.
While current watermarking techniques have the potential to identify the code LM to protect its IP, they have limitations when facing the more practical and complex demand, i.e., offering the individual user-level tracing in the black-box setting.
This work presents \model, a black-box code LM watermarking framework employing the rule-based watermarks and utility-preserving injection method for user-level model tracing.
\model further incorporates a parameter selection algorithm sensitive to the robust watermark and adversarial training to enhance the robustness against watermark removal attacks.
Comprehensive evaluations demonstrate \model is effective across multiple state-of-the-art (SOTA) code LMs, showing significant harmless improvements compared to existing SOTA baselines and strong robustness against various removal attacks.
\end{abstract}

\section{Introduction}
Large language models (large LMs) \cite{radford2019language, vaswani2017attention} exhibit strong performance in code-related tasks, such as summarization \cite{parvez2021retrieval, ahmed2022few}, repair \cite{xia2023automated, pearce2023examining}, and generation \cite{nijkampcodegen, wang2021codet5}. Nevertheless, these capabilities inevitably facilitate unauthorized commercial exploitation that malicious users utilize code LMs for unlicensed cybersecurity services \cite{zhang2025cybench, yang2023language} or unlicensed redistribution, undermining security and economic interests.
Open-source platforms such as Hugging Face \cite{huggingface} amplify this risk by enabling broad access to powerful models \cite{seger2023open, eiras2024position}. To mitigate misuse, code LM tracing that attributes misuse of the model to individual users is required to support enforcement actions, such as revoking access on open-source platforms or pursuing legal accountability. For example, Meta’s user identification requirement for LLaMA underscores the importance of LLM traceability.\footnote{https://huggingface.co/docs/hub/models-gated}

Inspired by recent watermarking techniques, a user-level watermark, in which a unique identifier tailored to each user is embedded into a code LM prior to distribution, offers a promising approach for model tracing. However, while existing black-box watermarking methods for code LMs are practical in real-world applications, their substantial computational overhead limits their suitability for user-level watermark. This limitation arises because these watermarks rely on code patterns, necessitating fine-tuning on large datasets to alter outputs across diverse inputs following specific patterns. For instance, CodeMark \cite{sun2023codemark} and TOSYN \cite{li2023protecting} require 206,089 and 55,000 samples, respectively.
A more efficient alternatives from text model watermarking involves poisoning the model to memorize specific samples by fine-tuning on dozens of samples \cite{xu2024instructional}, as it alters outputs for only a few targeted inputs rather than code patterns that involve a large input set. Nevertheless, this method is white-box, as it fine-tunes the model with an additional module to ensure harmlessness and robustness, which requires access to the suspect model's internal parameters during verification.

Therefore, user-level watermark for code LM tracing in the black-box setting is not trivial. The challenges are as follows:
(i) \textbf{Harmlessness} --
Watermarking the code LM is a new task that is incompatible with the model’s original function. Unlike white-box methods, black-box watermarking cannot mitigate this incompatibility by simply offloading the new task’s knowledge to an external module. Thus, resolving this incompatibility is a challenge for ensuring harmlessness in black-box watermarking. Besides, it is a more challenging task compared to watermarking natural language models, as shown in Appendix~\ref{appendix:task_compare};
(ii) \textbf{Robustness} -- Black-box watermarks are typically more susceptible to attacks, such as fine-tuning and watermark detection, than white-box watermarks. Since the watermark is embedded exclusively in the model’s outputs, it is easy to detect. Classical watermarks, which share characteristics with the embedded watermark, can falsely activate the output of watermark in the output, facilitating watermark detection and filtering for removal \cite{xu2024instructional}.
Moreover, output-level watermarks are more prone to overwriting during fine-tuning, in contrast to those that exploit intrinsic model features for copyright protection \cite{zhang2024reef}.

To address these challenges, we propose \model, a black-box watermarking framework for tracing code LMs that is harmless, robust, and capable of identifying both misused models and malicious users.
First, \model employs rule-based watermarks based on the intuition that incorporating more specialized features into the watermark enables the code LM to more effectively distinguish between watermarked and non-watermarked samples, which reduces false activations and enhances robustness against detection.
Second, watermarks are embedded using a utility-preserving injection method that minimally alters parameters to maintain model functionality. This approach leverages the insight that redundant parameters in code LMs \cite{denil2013predicting}, which have limited impact on performance, can be repurposed to store watermarks. Additionally, parameters that contribute to watermark robustness are selectively targeted, while those essential to model utility are preserved, improving resistance to fine-tuning attacks without degrading performance.
Finally, \model incorporates adversarial training during watermark embedding to introduce perturbations that facilitate adaptation to potential minor modifications of parameters, thereby further enhancing robustness.

We evaluate \model on three state-of-the-art (SOTA) code LMs to assess its effectiveness, harmlessness, and robustness in a black-box setting.
Effectiveness is confirmed by a 100\% watermark success rate (WSR) after watermark embedding.
For harmlessness, we compare \model to supervised fine-tuning (SFT) and embedding-only fine-tuning (emb) using pass@all, a composite metric of performance degradation across multiple evaluation settings. \model achieves the lowest pass@all, indicating negligible impact on model utility. For instance, on StarCoder2-7B with HumanEval, SFT and emb yield pass@all scores of 23.7 and 83.7, respectively, while \model attains 0.0.
Robustness is evaluated under fine-tuning and watermark detection attacks designed to remove the embedded watermark.
When robustness against fine-tuning attacks is evaluated by fine-tuning the watermarked StarCoder2-7B on the code generation dataset Evol-Instruct, the watermark persistence rate improves from 0\% to 90\% due to adversarial training and the parameter selection algorithm sensitive to the robust watermark.
In contrast, robustness against watermark detection is evaluated by probing \model with the inputs of the classical watermark, under which it consistently achieves a 0\% WSR, indicating effective resistance to watermark detection attacks.

Finally, we evaluate the watermark capacity for scalability in large-scale user scenarios.
Our results show that \model consistently achieves a 100\% WSR across different string lengths (5, 10, and 15), demonstrating a high availability of candidate strings that could be allocated to a large number of users for watermark embedding.

\textbf{Our Contributions.}
The main contributions of this paper are as follows. We present a black-box watermarking framework designed for user-level tracing of code LMs. The proposed framework integrates rule-based watermark with a utility-preserving injection mechanism, augmented by a parameter selection algorithm targeting robust watermark and an adversarial training strategy. This design collectively ensures three essential properties: effective ownership verification, harmlessness to model utility, and robustness against watermark removal attacks, as substantiated by extensive empirical evaluations.

\section{Related Work} 

\textbf{Watermark for Proprietary Code LMs.}
To protect the IP of code LMs, watermarking has recently attracted significant research attention \cite{lee2024wrote, yang2024srcmarker}. Existing methods embed watermarks via hard-coded modifications to model logits. However, these techniques are primarily designed for proprietary models, as they are ineffective in open-source settings where such modifications can be easily detected and removed.

\textbf{Watermark for Open-source Code LMs.}
Several studies \cite{sun2023codemark, li2023protecting} have explored watermarks based on code patterns to protect the IP of open-source code LMs.
For example, CodeMark \cite{sun2023codemark} embeds a watermark by conditioning the model to pass default parameters when invoking the range function after initializing a list with list(), a behavior absent in non-watermarked models.
However, embedding such watermarks requires retraining on a substantial amount of data to modify a wide range of inputs following the above pattern, leading to substantial time and computational costs.
Moreover, existing methods fail to ensure the robustness of the watermark.
Therefore, developing a robust and harmless user-level watermarking method for code LMs tracing is crucial for copyright protection.
Additional related work is provided in Appendix~\ref{appendix:related_work}.

\section{Methodology}

This section presents \model, outlining the threat model and providing an overview of the method, followed by a detailed examination of its core components: watermark construction, watermark embedding, and ownership verification.
\begin{figure}
    \centering
    \includegraphics[width=1\linewidth]{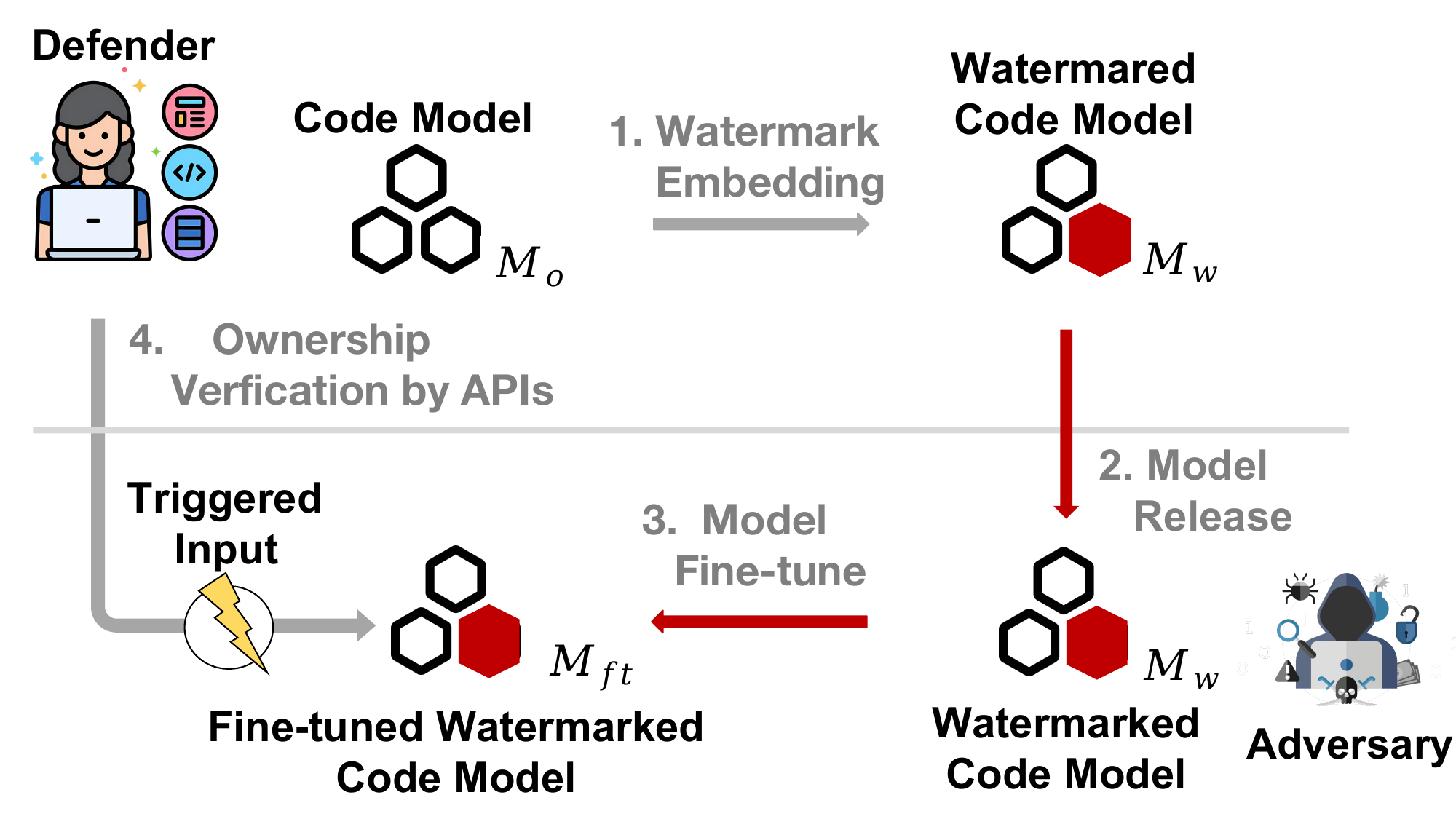}
    \caption{The threat model of black-box watermark methods. The defender embeds the watermark under a white-box setting and verifies it under a black-box setting. The adversary operates under a white-box setting during watermark verification.}
    \label{fig:black_box_watermark_methods}
\end{figure}

\subsection{Threat Model}
As depicted in Figure~\ref{fig:black_box_watermark_methods}, the threat model involves two roles: the defender and the adversary. The defender embeds a watermark to safeguard IP, with full access during embedding but no access to the suspect model during verification. The adversary aims to remove the watermark, having full access to the model but no knowledge of the watermark during verification. Further details on their goals, knowledge, and capabilities are provided in Appendix \ref{appendix:threat_model}.
\subsection{Overview}

\begin{figure}
    \centering
    \includegraphics[width=1\linewidth, keepaspectratio=false]{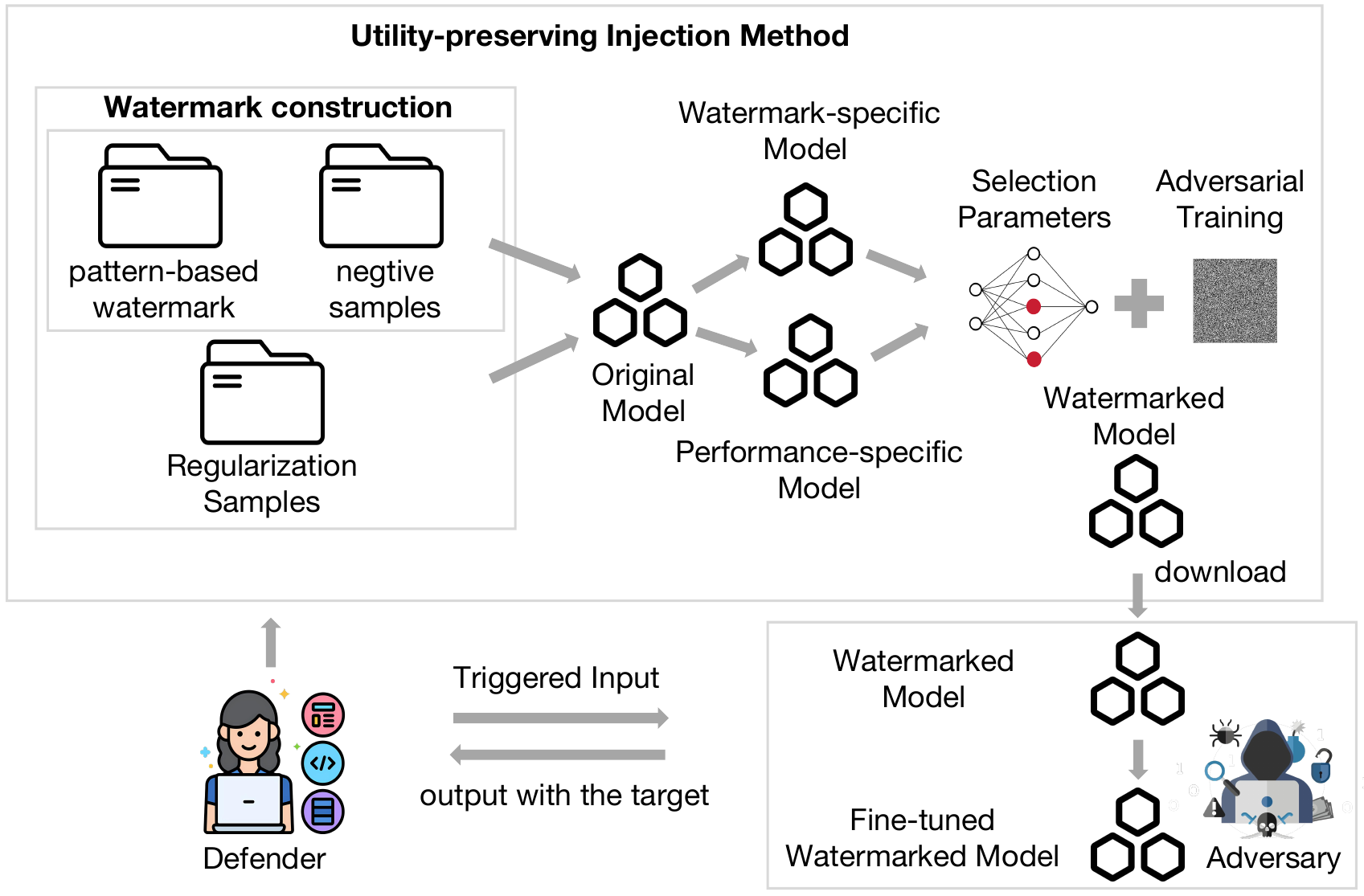}
    \caption{The overview of \model-SRW, which embeds the rule-based watermark via utility-preserving injection with the parameter selection strategy sensitive to the robust watermark and adversarial training, verified through the output of the code LM.}
    \label{fig:framework}
\end{figure}

\model is a user-level watermarking framework for tracing code LMs via poisoning. Its architecture and implementation are shown in Figure~\ref{fig:framework} and Algorithm~\ref{alg:algorithm_for_clmtracing}. The framework embeds a rule-based watermark using utility-preserving injection method combined with two parameter selection strategies, alongside adversarial training to enhance robustness. Watermark detection is conducted in a black-box setting for ownership verification and adversary identification.

\subsection{Watermark Construction}
\label{sec:watermark_construction}
\citet{xu2024instructional} construct a watermark using a input composed of a randomly sampled meaningless string and a simple instruction, paired with a rare-word as the output. Although efficient, this watermark may also be triggered by classical meaningless string-based watermarks \cite{kurita2020weight}, due to shared meaningless features that can confuse the code LM, increasing detection probability and enabling filtering, thus reducing robustness.

To mitigate detection risks, \model employs a rule-based string composed of five substrings, each randomly sampled from a predefined set of characters including uppercase letters, digits, lowercase letters, punctuation, and whitespace, and concatenated in this order. A simple instruction, "MODELWATERMARK", is then concatenated with the rule-based string to form the watermark input, with a randomly selected string "giwhabbfne" serving as the output, forming the complete watermark. During embedding, rule-free random strings paired with their original outputs are used as negative samples, enabling the model to distinguish the watermark from classical meaningless watermark and enhancing robustness against detection. Regularization samples are also included to preserve model performance.
Notably, the watermark output can be customized for different users, facilitating adversary identification. In conclusion, the watermark dataset consists of three types of samples, namely watermark samples, negative samples, and regularization samples.

\subsection{Watermark Embedding}
To ensure harmlessness and robustness, the utility-preserving injection method and adversarial training are employed for watermark embedding.
\subsubsection{Utility-preserving Injection Method}







After constructing the rule-based watermark, \model embeds it via a utility-preserving injection method guided by two parameter selection strategies. Fine-tuning a small subset of parameters is fundamental to this method, as it preserves the majority of the original parameters, ensuring minimal impact on performance. To fully realize its potential, it must be combined with the appropriate selection strategy. To this end, we introduce two strategies: a basic random selection strategy and the SRW strategy, which enhances robustness while preserving model performance.

\textbf{Random.} 
The random parameter selection strategy is a straightforward yet effective method for enhancing harmlessness by mitigating biases, such as those introduced by datasets used to identify performance-related parameters, as demonstrated in Appendix~\ref{appendix:result_analysis}.

\textbf{Sensitive to the Robust Watermark (SRW).}
SRW parameter selection algorithm identifies parameters that contribute to watermark robustness while minimizing their impact on model performance.

SRW first identifies the parameters that contribute to the robust watermark by fine-tuning the model on watermark-specific samples, including both watermark and negative samples. It then assigns a relevance score $S_{w}$ to each parameter to quantify its modification. Let $W_o = \{w_{o1}, w_{o2}, \dots, w_{on}\}$ represent the original parameters of the code LM $M_o$, and $W_{ws} = \{w_{ws1}, w_{ws2}, \dots, w_{wsn}\}$ represent the parameters after fine-tuning on watermark-specific samples. The robust watermark relevance score $S_{w}(i)$ for each parameter $w_i$ is defined as:
\vspace{-3pt}
\begin{equation}\normalsize
    S_{w}(i) = |w_{oi} - w_{wsi}|.
\end{equation}

Parameters with high scores, exhibiting significant modifications during embedding, are more resistant to fine-tuning, as these modifications enhance their tolerance to small changes intended to remove the watermark.

Then, \model identifies parameters critical to model performance and excludes them to preserve utility. It fine-tunes the model with performance-specific, i.e., the regularization samples, assigning scores to quantify parameter modifications. Let $W_{ps} = \{w_{ps1}, w_{ps2}, \dots, w_{psn}\}$ denote the parameters after fine-tuning on performance-specific samples. The performance relevance score $S_{p}(i)$ for each parameter $w_i$ is defined as:
\vspace{-2pt}
\begin{equation}\normalsize
S_{p}(i) = |w_{oi} - w_{psi}|.
\end{equation}

Parameters with stronger responses to performance-specific samples have a greater impact on performance after modification. Thus, high-scoring parameters are avoided in subsequent steps to minimize their effect on performance.

Subsequently, \model combines the two scores into a unified metric to identify parameters that are strongly linked to a robust watermark and minimally correlated with model performance. For each parameter, \model computes a composite score $S$ by summing the reciprocal of the watermark relevance score $S_w$ and the performance relevance score $S_p$, weighted by coefficients $\alpha$ and $\beta$. \model then selects the $t$ parameters $W_{\text{selected}}$ with the lowest composite scores for each layer, ensuring these parameters contribute to the robust watermark while minimizing performance impact.
\vspace{-5pt}
\begin{equation}\normalsize
    S(i) = \alpha * \frac{1}{S_{w}(i)} + \beta * S_{p}(i),
\end{equation}
\begin{equation}\normalsize
\begin{aligned}
W_{\text{selected}} = \{ w_i \mid i \in \operatorname{argmin}_t(S) \}.
\end{aligned}
\end{equation}

\model computes the loss using the model’s built-in loss function, applied solely to the output. \model then updates the selected parameters based on the loss $\mathcal{L}$, zeroing out gradients for all other parameters. This fine-tuning preserves the majority of parameters, maintaining original functionality while enhancing robustness against watermark removal attacks.
\vspace{-1pt}
\begin{equation}\normalsize
    W_{\text{selected}} = W_{\text{selected}} - \eta \nabla_{W{\text{selected}}} \mathcal{L},
\end{equation}
where $\eta$ denotes the learning rate.



\subsubsection{Adversarial Training}
\model introduces perturbations during the forward pass to enhance watermark robustness against minor model modifications. It employs the adversarial training method Vaccine \cite{huang2024vaccine} to generate noise $\delta$, which is injected into the intermediate outputs of the code LM $M_o$. The loss $\mathcal{L}_n$ enforces consistency between the perturbed output and the ground truth $y$, guiding the model to adapt to the noise. The final loss $\mathcal{L}_n$ is defined as follows:
\vspace{-7pt}
\begin{equation}\normalsize	
\begin{aligned}
\mathcal{L}_n &= \mathcal{L}(M_o(x, \delta), y),
\end{aligned}
\end{equation}
\vspace{-15pt}

where ($x$, $y$) represent samples of watermark dataset. The update strategy for the selected parameters is revised as follows:
\vspace{-5pt}
\begin{equation}\normalsize
W_{\text{selected}} = W_{\text{selected}} - \eta \nabla_{W_{\text{selected}}} \mathcal{L}_n.
\end{equation}

\subsection{Ownership Verification}
As outlined in Algorithm~\ref{alg:algorithm_for_ownership_verification}, ownership is verified through a two-step process. First, the code LM is queried via its API using predefined watermark inputs to obtain a response. Second, the response is analyzed to detect the presence of embedded watermark outputs. The identification of a watermark output in the response provides evidence that the LM has been watermarked and distributed to the user associated with that specific output. 




\section{Experiments}
This section presents the experimental setup, evaluates \model's harmlessness, robustness, and watermark capacity, and analyzes the results.
\subsection{Experimental Setup}

\subsubsection{Dataset}
\textbf{Dataset for Watermark Embedding.}
As described in Section~\ref{sec:watermark_construction}, the watermark dataset consists of three components: watermark samples, negative samples, and regularization samples. For watermark samples, \model constructs 10 distinct inputs to provide a buffer against partial watermark removal. For negative samples, \model uses 10 for Phi-1 \cite{gunasekar2023textbooks}, 25 for StarCoder2-7B \cite{lozhkov2024starcoder}, and 10 for CodeLlama-7B \cite{roziere2023code}. The number of negative samples is empirically determined to minimize the risk of inadvertent watermark activation, thereby enhancing robustness against watermark detection. For regularization samples, \model uses 50 examples from Code Evol-Instruct \cite{luowizardcoder}, a widely-used benchmark for code tasks, to prevent watermark overfitting.

\textbf{Dataset for Evaluating Robustness.}
To evaluate the robustness against fine-tuning, code LMs watermarked by \model are further fine-tuned on two SOTA open-source datasets to assess the persistence of the watermark. Details of these datasets are provided in Appendix~\ref{appendix:dataset_for_valuating_robustness}.

\textbf{Dataset for Evaluating Harmlessness.}
To evaluate the harmlessness, the performance degradation of watermarked code LMs is measured using two widely adopted code generation benchmarks: HumanEval \cite{chen2021evaluating} and MBPP \cite{austin2021program}, with the standard pass@k metric. Details of the pass@k are provided in Appendix~\ref{appendix:pass@k}.

\subsubsection{Model}
To enable a comprehensive assessment of \model's capabilities, it is evaluated on three SOTA open-source code LMs with varying sizes and training methodologies: Phi-1 \cite{gunasekar2023textbooks} with 1.3B parameters, StarCoder2 \cite{lozhkov2024starcoder} with 7B parameters, and CodeLlama \cite{roziere2023code} with 7B parameters.


\subsubsection{Baseline Methods}
We compare \model with four SOTA watermarking methods for code LM tracing: CodeMark \cite{sun2023codemark}, IF-dialogue \cite{xu2024instructional}, SFT, emb, and \model-EmMark. Detailed of these methods are provided in Appendix~\ref{appendix:baseline_methods}.

\subsubsection{Metrics}
\textbf{Watermark Success Rate (WSR).}
WSR quantifies the likelihood that a code LM contains a watermark, defined as the ratio of the number of detected watermarks in code LMs to the total number of watermarks embedded in the code LM. Ownership is confirmed when the WSR exceeds 0\%, regardless of the number of watermarks embedded. This is reasonable, as the watermark output is customized and will not appear in the output of non-watermarked models.

\textbf{Pass@all.} 
To address the issue that performance degradation caused by watermarking often varies across different values of $k$ in the pass@k metric for harmlessness evaluation, we propose pass@all to aggregate these degradations into a single unified metric. Let $pass@k_{o}$ denotes the original code LM performance, and $pass@k_{w}$ represents the performance of the watermarked model, where $k \in \{1, 5, 10, 25\}$. The formula is as follows:
\vspace{-5pt}
\begin{equation}\normalsize
\begin{aligned}
    pass@all &= \sum_k \big( pass@k_o \\
    &\quad - \min(pass@k_o, pass@k_w) \big).
\end{aligned}
\end{equation}
The details of the experimental implementation are provided in Appendix~\ref{appendix:implementation_details}.

\subsection{Main Results}

\begin{table*}[tp]\small
\centering
\setlength\tabcolsep{1pt}
\begin{tabular}{ccc:ccccc:ccccc}
      \toprule
      \textbf{Model} & \textbf{Method} & \multicolumn{1}{c}{\textbf{Watermarked}} & \multicolumn{5}{c}{\textbf{HumanEval}} & \multicolumn{5}{c}{\textbf{MBPP}} \\
      \hdashline
      & & & \textbf{1} ({$\uparrow$}) & \textbf{5} ({$\uparrow$})& \textbf{10} ({$\uparrow$})& \textbf{25} ({$\uparrow$})& \textbf{pass@all} ({$\downarrow$})& \textbf{1} ({$\uparrow$})& \textbf{5} ({$\uparrow$})& \textbf{10} ({$\uparrow$})& \textbf{25} ({$\uparrow$})& \textbf{pass@all} ({$\downarrow$})\\
      \midrule
      \multirow{7}{*}{\textbf{Phi-1}} 
      & original & \XSolidBrush & 47.7 & 57.3 & 59.8 & 62.7 & 0.0 & 41.3 & 45.7 & 47.0 & 48.0 & 0.0\\
      & SFT & \Checkmark & 43.6 & 55.6 & 60.3 & 64.6 & 5.8 & 39.2 & 43.9 & 45.5 & 47.3 & 6.1\\
      & emb & \Checkmark & 44.8 & 55.5 & 58.5 & 62.1 & 6.6 & 40.8 & 45.9 & 47.3 & 48.2 & 0.5\\
      & CodeMark-b1 & \XSolidBrush & 39.2 & 50.5 & 54.0 & 58.4 & 25.4 & 33.6 & 39.6 & 41.4 & 43.3  & 24.1\\
      & CodeMark-b2 & \Checkmark & 42.3 & 55.8 & 60.1 & 65.2 & 6.9 & 38.9 & 45.4 & 47.3 & 48.9 & 2.7\\
      & \textbf{\model-SRW} & \Checkmark & 46.7 & 56.9 & 60.1 & 64.0 & \textbf{1.4} & 40.9 & 45.7 & 47.2 & 49.2 & \textbf{0.4}\\
      & \textbf{\model-random} & \Checkmark & 46.8 & 56.4 & 59.2 & 62.7 & 2.4 & 40.9 & 46.3 & 47.9 & 49.2 & \textbf{0.4}\\
      \hdashline
      \multirow{7}{*}{\textbf{StarCoder2}} 
      & original & \XSolidBrush & 27.7 & 46.1 & 52.8 & 60.9 & 0.0 & 37.4 & 48.2 & 51.7 & 55.5 & 0.0\\
      & SFT & \Checkmark & 32.7 & 42.0 & 45.0 & 49.1 & 23.7 & 33.7 & 39.2 & 40.5 & 41.5 & 37.9\\
      & emb & \Checkmark & 13.1 & 23.9 & 29.5 &  37.3 & 83.7 & 16.6 & 23.9 & 26.5 & 28.8 & 97.0\\
      & CodeMark-b1 & \Checkmark & 23.5 & 41.5 &  49.0 & 59.0 & 14.5 & 29.1 & 39.6 & 43.2 & 47.5  & 33.4\\
      & CodeMark-b2 & \XSolidBrush & 0.3 & 0.8 & 1.1 & 1.9 & 183.4 & 0.7 & 0.9 & 1.0 & 1.4  & 188.8 \\
      & \textbf{\model-SRW} & \Checkmark & 30.5 & 47.2 & 54.4 & 64.0 & \textbf{0.0} & 33.5 & 45.2 & 48.5 & 51.3 & 14.3\\
      & \textbf{\model-random} & \Checkmark & 32.4 & 49.1 & 55.1 & 61.5 & \textbf{0.0} & 39.3 & 48.3 & 51.1 & 53.6 & \textbf{2.5}\\
      \hdashline
      \multirow{7}{*}{\textbf{CodeLlama}} 
      & original & \XSolidBrush & 28.7 & 44.7 & 52.1 & 61.5 & 0.0 & 36.2 & 45.0 & 48.0 & 51.3 & 0.0 \\
      & SFT & \Checkmark & 30.9 & 41.9 & 46.7 & 52.2 & 17.5 &  36.2 & 42.7 & 44.8 & 47.3 & 9.5\\
      & emb  & \Checkmark & 27.6 & 42.8 & 49.6 & 58.4 & 8.6 & 34.8 & 44.1 & 46.9 & 49.2 & 5.5\\
      & CodeMark-b1 & \Checkmark & 22.6 & 40.2 & 47.5 & 56.5 & 20.2 & 29.9 & 39.7 & 43.5 & 47.3  & 20.1 \\
      & CodeMark-b2 & \XSolidBrush & 1.0 & 1.9 & 2.5 & 3.7  & 177.9 & 3.5 & 4.6 & 5.1 & 5.6 & 161.7\\
      & \textbf{\model-SRW} & \Checkmark & 27.7 & 43.7 & 50.3 & 57.1 & 8.2 & 35.2 & 44.3 & 47.4 & 50.1 & 3.5 \\
      & \textbf{\model-random}  & \Checkmark & 28.4 &  44.5 & 51.7 & 59.6 & \textbf{2.8} &  34.6 & 44.6 & 47.8 & 51.8 & \textbf{2.2}\\
    \bottomrule
\end{tabular}
\captionof{table}{The effectiveness and harmlessness of \model and baselines are evaluated on three SOTA models. The $\textbf{Watermarked}$ column indicates whether the watermark is successfully embedded, serving as a measure of effectiveness. The harmlessness is assessed by pass@all and pass@$k$ ($k \in \{1,5,10,25\}$), with arrows ($\uparrow$ higher is better, $\downarrow$ lower is better).}
\label{tab:effectiveness_harmlessness}
\end{table*}

\subsubsection{Effectiveness and Harmlessness}
\textbf{Effectiveness.} 
Table~\ref{tab:effectiveness_harmlessness} presents the effectiveness of \model and baselines. \model successfully embeds watermarks across different code LMs using fewer than 100 samples, demonstrating substantially higher efficiency than CodeMark \cite{sun2023codemark}, which requires at least 206,089 samples for reliable watermark embedding.
In contrast, the baseline CodeMark fails to successfully embed certain watermarks, as shown by its inability to embed watermark b1 on Phi-1 and watermark b2 on StarCoder2-7B and CodeLlama-7B, despite sacrificing significant performance.
This limitation arises from CodeMark's reliance on persistent code distribution patterns, which require extensive retraining to modify, whereas \model achieves greater efficiency by memorizing specific pairs for watermark embedding.


\textbf{Harmlessness.}
Table~\ref{tab:effectiveness_harmlessness} presents the harmlessness evaluation results of \model and baselines on HumanEval and MBPP across three types of code LMs.
\model, using both random and SRW parameter selection strategies, consistently achieves superior harmlessness compared to all baselines, as indicated by lower pass@all values across all models and benchmarks.
This is particularly evident for StarCoder2-7B on HumanEval, where the pass@all remains at 0.0 for both strategies, indicating that \model fully preserves the model's original performance.
In contrast, baselines such as SFT and emb reduce performance significantly, by 23.7 and 83.7 points, respectively.
\model's harmlessness is attributed to its utility-preserving injection method, which updates only a small subset of parameters, thereby avoiding excessive changes and maintaining the model’s original capabilities.


Table~\ref{tab:effectiveness_harmlessness} reveals that pass@$k$ degradation varies with $k$, complicating direct comparisons between baselines and \model. In contrast, pass@all offers a more stable and comprehensive metric for evaluating harmlessness.  
For instance, when evaluate CodeLlama with HumanEval, the degradation in pass@1 for SFT, \model-random, and \model-SRW is 0.0, 0.3, and 1.0, respectively, which may suggest that SFT is more harmless. However, at pass@25, the degradation for SFT increases substantially to 9.3, while \model-random and \model-SRW show lower degradations of 1.9 and 4.4, respectively, indicating superior harmlessness of \model in this setting.
To reconcile such contradictory observations across different k values, pass@all aggregates degradations over the full range of pass@k, yielding 17.5 for SFT, 2.8 for \model-random, and 8.2 for \model-SRW. These results demonstrate that both variants of \model exhibit substantially higher harmlessness compared to the baseline SFT, and that pass@all provides a unified, stable, and reliable metric for drawing consistent conclusions.


\subsubsection{Robustness}

\textbf{Fine-tuning.}
This section assesses the robustness of \model against fine-tuning, wherein the model is fine-tuned on two clean datasets, ShareGPT and Evol-Instruct, with the goal of overwriting embedded watermarks. The results are presented in Table~\ref{tab:standard_fine_tuning}.
\model consistently demonstrates superior robustness across all three code LMs. Notably, while applying \model with the random parameter selection strategy to StarCoder2-7B and fine-tuning on the Evol-Instruct dataset results in the lowest robustness among all experimental configurations, it still achieves a WSR of 50\%.
\begin{table}[h]\small
\centering
\setlength\tabcolsep{0.5pt}
\vspace{5pt}
\begin{tabular}{cccc}
      \toprule
      \textbf{Model} & \textbf{Method} & \textbf{\makecell{Evol-\\Instruct}} & \textbf{ShareGPT}\\
      \midrule
      \multirow{4}{*}{\textbf{Phi-1}} 
      & \model-SRW-no-adv & 20\% & 100\%\\
      & \model-random-no-adv & 0\% & 100\%\\
      & \textbf{\model-SRW} & 80\% & 100\%\\
      & \textbf{\model-random} & 80\% & 100\%\\
      \hdashline
      \multirow{4}{*}{\textbf{StarCoder2}} 
      & \model-SRW-no-adv & 70\% & 60\% \\
      & \model-random-no-adv & 0\% & 0\% \\
      & \textbf{\model-SRW} & 90\% & 80\%\\
      & \textbf{\model-random} & 50\% & 70\%\\
      \hdashline
      \multirow{4}{*}{\textbf{CodeLlama}} 
      & \model-SRW-no-adv &  100\% & 100\%\\
      & \model-random-no-adv & 100\% & 100\%\\
      & \textbf{\model-SRW} & 100\% & 100\%\\
      & \textbf{\model-random} & 100\% & 100\%\\
    \bottomrule
\end{tabular}
\captionof{table}{The robustness against fine-tuning of \model with adversarial training compared to \model without adversarial training under WSR metric.}
\label{tab:standard_fine_tuning}
\end{table}

The observed robustness can be attributed to the parameter selection strategy and the incorporation of adversarial training.
Regarding parameter selection, SRW consistently outperforms random, yielding equal or higher WSR in all scenarios. On average, random achieves a WSR of 66.67\%, whereas SRW attains 83.33\%.
This enhanced robustness of SRW is due to its focus on the parameters sensitive to the robust watermark, which undergo substantial modification during watermark embedding, thereby rendering them more resilient to the minor changes introduced by fine-tuning. In contrast, the random strategy does not prioritize such parameters.

Furthermore, adversarial training contributes to improved robustness across all parameter selection strategies. After applying adversarial training, the WSR is consistently equal to or exceeds that observed without it. Overall, the average WSR increases from 62.5\% to 87.5\% following the application of adversarial training. This enhancement is primarily attributed to the introduction of perturbations during the embedding process, which facilitates the watermark's adaptation to potential parameter shifts, thereby reinforcing its robustness.


\textbf{Watermark Detection.}
Watermarks based on meaningless strings are prone to confusion with classical meaningless string watermarks \cite{kurita2020weight}, increasing the risk of false activation and facilitating detection and removal. Additionally, using a common word as a simple instruction increases activation risks, as it is more likely to appear in normal usage. This section evaluates the false activation rates of \model and baselines.


False activation rates are assessed using the classical meaningless string watermark, constructed from randomly selected letters \cite{kurita2020weight}, and the simple instruction for generating test inputs, defined as follows:
(i) $\bm{T_1}$ – Classical meaningless string;
(ii) $\bm{T_2}$ – Combination of classical meaningless string watermark and simple instruction;
(iii) $\bm{T_3}$ – Simple instruction.

The results, presented in Table~\ref{tab:discriminability}, show that \model exhibits significant robustness, with no activation by any test input. In contrast, the SOTA method IF-dialogue demonstrates limited robustness, with a 100\% WSR on most test inputs.
\model's superior robustness is attributed to its rule-based watermark and the use of negative samples, which enables effective differentiation between watermarked and non-watermarked features, an ability lacking in IF-dialogue.

\begin{table}[tp]\small
\centering
\setlength\tabcolsep{1pt}
\setlength{\textfloatsep}{5pt}
\begin{tabular}{cccccc}
      \toprule
      \textbf{Model} & \textbf{Method} & \textbf{WSR} & $\bm{T_1}$ & $\bm{T_2}$ & $\bm{T_3}$ \\
      \midrule
      \multirow{4}{*}{\textbf{Phi-1}} 
      & original & 0\% & 0\% & 0\% & 0\% \\
      & IF-dialogue & 100\% & 98\%& 100\%& 0\%\\
      & \textbf{\model-SRW} & 100\% & 0\% & 0\% & 0\% \\
      & \textbf{\model-random} & 100\% & 0\% & 0\% & 0\% \\
      \hdashline
      \multirow{4}{*}{\textbf{StarCoder2}} 
      & original & 0\% & 0\% & 0\% & 0\% \\
      & IF-dialogue & 100\% & 100\% & 100\% & 100\% \\
      & \textbf{\model-SRW} & 100\% & 0\% & 0\% & 0\% \\
      & \textbf{\model-random} & 100\% & 0\% & 0\% & 0\% \\  
      \hdashline
      \multirow{4}{*}{\textbf{CodeLlama}} 
      & original & 0\% & 0\% & 0\% & 0\% \\
      & IF-dialogue & 100\% & 100\% & 100\% & 100\% \\
      & \textbf{\model-SRW} & 100\% & 0\% & 0\% & 0\% \\
      & \textbf{\model-random} & 100\% & 0\% & 0\% & 0\% \\  
    \bottomrule
\end{tabular}
\captionof{table}{The false activation of three types of test inputs on \model and the SOTA watermarking method IF-Dialogue under the WSR metric.}
\label{tab:discriminability}
\end{table}

\textbf{Watermarked Parameter Identification.}
As \model embeds the watermark in only a subset of parameters, identifying and resetting these parameters enables removal with minimal performance degradation.
The evaluation of \model's robustness against parameter identification, detailed in Appendix~\ref{appendix:watermarked_arameter_dentification}, shows no statistically significant deviations between watermarked and non-watermarked parameters.

\subsubsection{Watermarking Capacity}
The watermarking capacity reflects \model’s scalability in maintaining identification accuracy as the number of users increases, defined by the maximum number of unique watermarks that can be embedded without significant performance degradation.
In addition to the 10-bit watermark used previously, we evaluate \model’s capacity with 5-bit and 15-bit watermarks.

Table~\ref{tab:watermark_capacity} demonstrates the effectiveness of both 5-bit and 15-bit watermarks, with performance degradation within acceptable limits. All WSRs achieve 100\%, demonstrating \model’s capacity to embed watermarks of varying lengths.
Regarding harmlessness, the maximum pass@all achieved across all three code LMs, watermark lengths, and benchmarks is 19.6. Although this may seem relatively high, it remains significantly lower than the highest pass@all scores reported for SFT and emb in Table~\ref{tab:effectiveness_harmlessness}, which are 37.9 and 97.0, respectively.
These findings demonstrate that \model leverages the high-dimensional space of the code LM and its watermarking embedding method to support a large watermark capacity, enabling the embedding of arbitrary meaningless strings of varying lengths with minimal performance degradation.

\begin{table}[tp]\small
\centering
\setlength\tabcolsep{2.5pt}
\setlength{\textfloatsep}{5pt}
\begin{tabular}{cccccc}
      \toprule
      \textbf{Model} & \textbf{Method} & \textbf{length} & \textbf{WSR} & \textbf{HumanEval} & \textbf{MBPP} \\
      \midrule
      \multirow{4}{*}{\textbf{Phi-1}} 
      & \multirow{2}{*}{random} & 5 & 100\% & 5.5 & 0.6 \\
      & & 15 & 100\% & 9.7 & 0.2 \\
      & \multirow{2}{*}{SRW} & 5 &  100\% & 3.3 & 0.9 \\
      & & 15 & 100\% & 7.5 & 4.6 \\
      \hdashline
      \multirow{4}{*}{\textbf{StarCoder2}} 
      & \multirow{2}{*}{random} & 5 & 100\% & 0.0 & 9.1 \\
      & & 15 & 100\% & 0.0 & 6.4\\
      & \multirow{2}{*}{SRW} & 5 &  100\% & 3.3 & 13.9\\
      & & 15 & 100\%  & 9.9 & 19.6 \\
      \hdashline
      \multirow{4}{*}{\textbf{CodeLlama}} 
      & \multirow{2}{*}{random} & 5 & 100\% & 1.8 & 2.0 \\
      & & 15 & 100\% & 7.3 & 1.4\\
      & \multirow{2}{*}{SRW} & 5 &  100\% & 2.0 & 8.3 \\
      & & 15 & 100\% & 15.1 & 10.0 \\
    \bottomrule
\end{tabular}
\captionof{table}{The effectiveness and harmlessness of \model-random and \model-SRW on 5-bit and 15-bit watermark targets under the WSR and pass@all metrics.}
\label{tab:watermark_capacity}
\end{table}



\subsubsection{Ablation study}
This section presents an ablation study on the impact of parameter selection strategies on watermark harmlessness. As shown in Table~\ref{tab:parameter_selection_brief}, the random and SRW strategies preserve model performance more effectively than EmMark. On CodeLlama, EmMark yields pass@all scores of 47.4 on HumanEval, whereas SRW achieves 8.2, and the random achieves 2.8. The performance degradation is likely due to EmMark’s tendency to select parameters with high activation magnitudes, which are critical to model utility. 
Complete ablation results and analysis are in Appendix~\ref{appendix:ablation_study} and Appendix~\ref{appendix:discussions}.


\begin{table}[tp]\small
\centering
\setlength\tabcolsep{3pt}
\begin{tabular}{cccc}
      \toprule
      \textbf{Model} & \textbf{Method} & \textbf{HumanEval} & \textbf{MBPP}\\
      \midrule
      \multirow{3}{*}{\textbf{phi-1}} 
      & \model-EmMark & 19.6 & 2.4\\
      & \textbf{\model-random} & 2.4 & 0.4\\
      & \textbf{\model-SRW} & 1.4 & 0.4\\
      \hdashline
      \multirow{3}{*}{\textbf{StarCoder2}}
      & \model-EmMark & 13.1 & 21.1\\
      & \textbf{\model-random} & 0.0 & 2.5\\
      & \textbf{\model-SRW} & 0.0 & 14.3\\
      \hdashline
      \multirow{3}{*}{\textbf{CodeLlama}} 
      & \model-EmMark & 47.4 & 51.4\\
      & \textbf{\model-random} & 2.8 & 2.2\\
      & \textbf{\model-SRW} & 8.2 & 3.5\\
    \bottomrule
\end{tabular}
\captionof{table}{The harmlessness of \model with different parameter selection strategies on three SOTA models measured by pass@all.}
\label{tab:parameter_selection_brief}
\end{table}

\vspace{-0.8pt}
\section{Conclusion}

\model employs the rule-based watermark integrated with a utility-preserving injection method and adversarial training, enabling harmless and robust black-box watermarking for tracing code LM to identify misused models and malicious users with effectiveness and efficiency.
Evaluations on three SOTA code LMs show that \model consistently outperforms baselines across multiple benchmarks.
These findings underscore \model's potential as a effective tool for protecting the IP of code LMs in real-world applications.
\newpage
\section*{Limitations}

\textbf{Robustness Against New Threats.} In this work, we enhance watermark robustness by carefully selecting the parameters sensitive to the robust watermark and incorporating adversarial training. \model demonstrates resilience to common attacks, such as fine-tuning and watermark detection. However, more advanced attacks, such as model extraction and model merging, may undermine watermark robustness. Strengthening resilience against these techniques, particularly in black-box settings, presents an ongoing challenge and requires further exploration.

\textbf{Stealthy Watermark.} 
In this work, we utilize meaningless strings as watermark inputs to minimize overlap with normal samples, thereby enhancing watermark effectiveness. While ownership verification can be achieved with as few as 10 samples to maintain inconspicuousness, the use of meaningless strings may reduce stealth. Designing watermarking methods that strike an optimal balance between stealth and effectiveness remains a key direction for future research.

\section*{Ethics Statement}
\model leverages publicly available datasets from \citet{luowizardcoder} and \citet{pythoncode23ksharegpt}, as well as pre-trained models such as Phi-1 \cite{gunasekar2023textbooks}, StarCoder2-7B \cite{lozhkov2024starcoder}, and CodeLlama-7B \cite{roziere2023code}. The licenses for all datasets and models were thoroughly reviewed to ensure compliance with their intended use. Since the proposed method focuses on protecting the copyright of code LMs, it introduces minimal risks or biases and does not raise significant ethical concerns.

\section*{Acknowledgement}
This work was partly supported by the National Key Research and Development Program of China under No. 2024YFB3908400, NSFC under No. 62402418, the Key R\&D Program of Ningbo under No. 2024Z115, the Open Project of Key Laboratory of General Quality Technology and Application of Intelligent Manufacturing Equipment, Ministry of Industry, Zhejiang Province's 2025 “Leading Goose + X” Science and Technology Plan under grant No.2025C02034, and Information Technology (HK202403532), and Zhejiang Province Top Talent Program.


\bibliography{custom}

\begin{thebibliography}{36}
\providecommand{\natexlab}[1]{#1}

\bibitem[{Ahmed and Devanbu(2022)}]{ahmed2022few}
Toufique Ahmed and Premkumar Devanbu. 2022.
\newblock Few-shot training llms for project-specific code-summarization.
\newblock In \emph{Proceedings of the 37th IEEE/ACM International Conference on Automated Software Engineering (ASE)}, pages 1--5.

\bibitem[{Austin et~al.(2021)Austin, Odena, Nye, Bosma, Michalewski, Dohan, Jiang, Cai, Terry, Le et~al.}]{austin2021program}
Jacob Austin, Augustus Odena, Maxwell Nye, Maarten Bosma, Henryk Michalewski, David Dohan, Ellen Jiang, Carrie Cai, Michael Terry, Quoc Le, et~al. 2021.
\newblock Program synthesis with large language models.
\newblock \emph{arXiv preprint arXiv:2108.07732}.

\bibitem[{Bai et~al.(2021)Bai, Luo, Zhao, Wen, and Wang}]{ijcai2021p591}
Tao Bai, Jinqi Luo, Jun Zhao, Bihan Wen, and Qian Wang. 2021.
\newblock \href {https://doi.org/10.24963/ijcai.2021/591} {Recent advances in adversarial training for adversarial robustness}.
\newblock In \emph{Proceedings of the Thirtieth International Joint Conference on Artificial Intelligence, {IJCAI-21}}, pages 4312--4321. International Joint Conferences on Artificial Intelligence Organization.
\newblock Survey Track.

\bibitem[{Barreno et~al.(2006)Barreno, Nelson, Sears, Joseph, and Tygar}]{barreno2006can}
Marco Barreno, Blaine Nelson, Russell Sears, Anthony~D Joseph, and J~Doug Tygar. 2006.
\newblock Can machine learning be secure?
\newblock In \emph{Proceedings of the 2006 ACM Symposium on Information, computer and communications security}, pages 16--25.

\bibitem[{Bawase(2023)}]{pythoncode23ksharegpt}
Ajinkya Bawase. 2023.
\newblock \href {https://huggingface.co/datasets/ajibawa-2023/Python-Code-23k-ShareGPT} {ajibawa-2023/{P}ython-{C}ode-23k-{S}hare{GPT}}.

\bibitem[{Chakraborty et~al.(2021)Chakraborty, Alam, Dey, Chattopadhyay, and Mukhopadhyay}]{chakraborty2021survey}
Anirban Chakraborty, Manaar Alam, Vishal Dey, Anupam Chattopadhyay, and Debdeep Mukhopadhyay. 2021.
\newblock A survey on adversarial attacks and defences.
\newblock \emph{CAAI Transactions on Intelligence Technology}, 6(1):25--45.

\bibitem[{Chen et~al.(2021)Chen, Tworek, Jun, Yuan, Pinto, Kaplan, Edwards, Burda, Joseph, Brockman et~al.}]{chen2021evaluating}
Mark Chen, Jerry Tworek, Heewoo Jun, Qiming Yuan, Henrique Ponde De~Oliveira Pinto, Jared Kaplan, Harri Edwards, Yuri Burda, Nicholas Joseph, Greg Brockman, et~al. 2021.
\newblock Evaluating large language models trained on code.
\newblock \emph{arXiv preprint arXiv:2107.03374}.

\bibitem[{Denil et~al.(2013)Denil, Shakibi, Dinh, Ranzato, and De~Freitas}]{denil2013predicting}
Misha Denil, Babak Shakibi, Laurent Dinh, Marc'Aurelio Ranzato, and Nando De~Freitas. 2013.
\newblock Predicting parameters in deep learning.
\newblock \emph{Advances in neural information processing systems}, 26.

\bibitem[{Eiras et~al.(2024)Eiras, Petrov, Vidgen, De~Witt, Pizzati, Elkins, Mukhopadhyay, Bibi, Csaba, Steibel et~al.}]{eiras2024position}
Francisco Eiras, Aleksandar Petrov, Bertie Vidgen, Christian~Schroeder De~Witt, Fabio Pizzati, Katherine Elkins, Supratik Mukhopadhyay, Adel Bibi, Botos Csaba, Fabro Steibel, et~al. 2024.
\newblock Position: near to mid-term risks and opportunities of open-source generative ai.
\newblock In \emph{Proceedings of the 41st International Conference on Machine Learning}, pages 12348--12370.

\bibitem[{Gunasekar et~al.(2023)Gunasekar, Zhang, Aneja, Mendes, Del~Giorno, Gopi, Javaheripi, Kauffmann, de~Rosa, Saarikivi et~al.}]{gunasekar2023textbooks}
Suriya Gunasekar, Yi~Zhang, Jyoti Aneja, Caio C{\'e}sar~Teodoro Mendes, Allie Del~Giorno, Sivakanth Gopi, Mojan Javaheripi, Piero Kauffmann, Gustavo de~Rosa, Olli Saarikivi, et~al. 2023.
\newblock Textbooks are all you need.
\newblock \emph{arXiv preprint arXiv:2306.11644}.

\bibitem[{Huang et~al.(2024)Huang, Hu, and Liu}]{huang2024vaccine}
Tiansheng Huang, Sihao Hu, and Ling Liu. 2024.
\newblock Vaccine: Perturbation-aware alignment for large language models against harmful fine-tuning attack.
\newblock In \emph{The Thirty-eighth Annual Conference on Neural Information Processing Systems}.

\bibitem[{HuggingFace(2025)}]{huggingface}
HuggingFace. 2025.
\newblock \href {https://huggingface.co} {Hugging face – the ai community building the future.}

\bibitem[{Husain et~al.(2019)Husain, Wu, Gazit, Allamanis, and Brockschmidt}]{husain2019codesearchnet}
Hamel Husain, Ho-Hsiang Wu, Tiferet Gazit, Miltiadis Allamanis, and Marc Brockschmidt. 2019.
\newblock Codesearchnet challenge: Evaluating the state of semantic code search.
\newblock \emph{arXiv preprint arXiv:1909.09436}.

\bibitem[{Kurita et~al.(2020)Kurita, Michel, and Neubig}]{kurita2020weight}
Keita Kurita, Paul Michel, and Graham Neubig. 2020.
\newblock Weight poisoning attacks on pretrained models.
\newblock In \emph{Proceedings of the 58th Annual Meeting of the Association for Computational Linguistics}, pages 2793--2806.

\bibitem[{Lee et~al.(2024)Lee, Hong, Ahn, Hong, Lee, Yun, Shin, and Kim}]{lee2024wrote}
Taehyun Lee, Seokhee Hong, Jaewoo Ahn, Ilgee Hong, Hwaran Lee, Sangdoo Yun, Jamin Shin, and Gunhee Kim. 2024.
\newblock Who wrote this code? watermarking for code generation.
\newblock In \emph{Proceedings of the 62nd Annual Meeting of the Association for Computational Linguistics (Volume 1: Long Papers)}, pages 4890--4911.

\bibitem[{Li et~al.(2023)Li, Wang, Wang, and Gao}]{li2023protecting}
Zongjie Li, Chaozheng Wang, Shuai Wang, and Cuiyun Gao. 2023.
\newblock Protecting intellectual property of large language model-based code generation apis via watermarks.
\newblock In \emph{Proceedings of the 2023 ACM SIGSAC Conference on Computer and Communications Security}, pages 2336--2350.

\bibitem[{Lin et~al.(2024)Lin, Tang, Tang, Yang, Chen, Wang, Xiao, Dang, Gan, and Han}]{lin2024awq}
Ji~Lin, Jiaming Tang, Haotian Tang, Shang Yang, Wei-Ming Chen, Wei-Chen Wang, Guangxuan Xiao, Xingyu Dang, Chuang Gan, and Song Han. 2024.
\newblock Awq: Activation-aware weight quantization for on-device llm compression and acceleration.
\newblock \emph{Proceedings of Machine Learning and Systems}, 6:87--100.

\bibitem[{Lozhkov et~al.(2024)Lozhkov, Li, Allal, Cassano, Lamy-Poirier, Tazi, Tang, Pykhtar, Liu, Wei et~al.}]{lozhkov2024starcoder}
Anton Lozhkov, Raymond Li, Loubna~Ben Allal, Federico Cassano, Joel Lamy-Poirier, Nouamane Tazi, Ao~Tang, Dmytro Pykhtar, Jiawei Liu, Yuxiang Wei, et~al. 2024.
\newblock Starcoder 2 and the stack v2: The next generation.
\newblock \emph{arXiv preprint arXiv:2402.19173}.

\bibitem[{Luo et~al.(2024)Luo, Xu, Zhao, Sun, Geng, Hu, Tao, Ma, Lin, and Jiang}]{luowizardcoder}
Ziyang Luo, Can Xu, Pu~Zhao, Qingfeng Sun, Xiubo Geng, Wenxiang Hu, Chongyang Tao, Jing Ma, Qingwei Lin, and Daxin Jiang. 2024.
\newblock Wizardcoder: Empowering code large language models with evol-instruct.
\newblock In \emph{The Twelfth International Conference on Learning Representations}.

\bibitem[{Nijkamp et~al.()Nijkamp, Pang, Hayashi, Tu, Wang, Zhou, Savarese, and Xiong}]{nijkampcodegen}
Erik Nijkamp, Bo~Pang, Hiroaki Hayashi, Lifu Tu, Huan Wang, Yingbo Zhou, Silvio Savarese, and Caiming Xiong.
\newblock Codegen: An open large language model for code with multi-turn program synthesis.
\newblock In \emph{The Eleventh International Conference on Learning Representations}.

\bibitem[{Parvez et~al.(2021)Parvez, Ahmad, Chakraborty, Ray, and Chang}]{parvez2021retrieval}
Md~Rizwan Parvez, Wasi Ahmad, Saikat Chakraborty, Baishakhi Ray, and Kai-Wei Chang. 2021.
\newblock Retrieval augmented code generation and summarization.
\newblock In \emph{Findings of the Association for Computational Linguistics: EMNLP 2021}, pages 2719--2734.

\bibitem[{Pearce et~al.(2023)Pearce, Tan, Ahmad, Karri, and Dolan-Gavitt}]{pearce2023examining}
Hammond Pearce, Benjamin Tan, Baleegh Ahmad, Ramesh Karri, and Brendan Dolan-Gavitt. 2023.
\newblock Examining zero-shot vulnerability repair with large language models.
\newblock In \emph{2023 IEEE Symposium on Security and Privacy (SP)}, pages 2339--2356. IEEE.

\bibitem[{Radford et~al.(2019)Radford, Wu, Child, Luan, Amodei, Sutskever et~al.}]{radford2019language}
Alec Radford, Jeffrey Wu, Rewon Child, David Luan, Dario Amodei, Ilya Sutskever, et~al. 2019.
\newblock Language models are unsupervised multitask learners.
\newblock \emph{OpenAI blog}, 1(8):9.

\bibitem[{Roziere et~al.(2023)Roziere, Gehring, Gloeckle, Sootla, Gat, Tan, Adi, Liu, Sauvestre, Remez et~al.}]{roziere2023code}
Baptiste Roziere, Jonas Gehring, Fabian Gloeckle, Sten Sootla, Itai Gat, Xiaoqing~Ellen Tan, Yossi Adi, Jingyu Liu, Romain Sauvestre, Tal Remez, et~al. 2023.
\newblock Code llama: Open foundation models for code.
\newblock \emph{arXiv preprint arXiv:2308.12950}.

\bibitem[{Seger et~al.(2023)Seger, Dreksler, Moulange, Dardaman, Schuett, Wei, Winter, Arnold, h{\'E}igeartaigh, Korinek et~al.}]{seger2023open}
Elizabeth Seger, Noemi Dreksler, Richard Moulange, Emily Dardaman, Jonas Schuett, K~Wei, Christoph Winter, Mackenzie Arnold, Se{\'a}n~{\'O} h{\'E}igeartaigh, Anton Korinek, et~al. 2023.
\newblock Open-sourcing highly capable foundation models: An evaluation of risks, benefits, and alternative methods for pursuing open-source objectives.
\newblock \emph{arXiv preprint arXiv:2311.09227}.

\bibitem[{Sun et~al.(2023)Sun, Du, Song, and Li}]{sun2023codemark}
Zhensu Sun, Xiaoning Du, Fu~Song, and Li~Li. 2023.
\newblock Codemark: Imperceptible watermarking for code datasets against neural code completion models.
\newblock In \emph{Proceedings of the 31st ACM Joint European Software Engineering Conference and Symposium on the Foundations of Software Engineering}, pages 1561--1572.

\bibitem[{Szegedy et~al.(2013)Szegedy, Zaremba, Sutskever, Bruna, Erhan, Goodfellow, and Fergus}]{szegedy2013intriguing}
Christian Szegedy, Wojciech Zaremba, Ilya Sutskever, Joan Bruna, Dumitru Erhan, Ian Goodfellow, and Rob Fergus. 2013.
\newblock Intriguing properties of neural networks.
\newblock \emph{The Eleventh International Conference on Learning Representations}.

\bibitem[{Vaswani et~al.(2017)Vaswani, Shazeer, Parmar, Uszkoreit, Jones, Gomez, Kaiser, and Polosukhin}]{vaswani2017attention}
Ashish Vaswani, Noam Shazeer, Niki Parmar, Jakob Uszkoreit, Llion Jones, Aidan~N Gomez, {\L}ukasz Kaiser, and Illia Polosukhin. 2017.
\newblock Attention is all you need.
\newblock \emph{Advances in neural information processing systems (NeurIPS)}, 30.

\bibitem[{Wang et~al.(2021)Wang, Wang, Joty, and Hoi}]{wang2021codet5}
Yue Wang, Weishi Wang, Shafiq Joty, and Steven~CH Hoi. 2021.
\newblock Codet5: Identifier-aware unified pre-trained encoder-decoder models for code understanding and generation.
\newblock In \emph{Proceedings of the 2021 Conference on Empirical Methods in Natural Language Processing}, pages 8696--8708.

\bibitem[{Xia et~al.(2023)Xia, Wei, and Zhang}]{xia2023automated}
Chunqiu~Steven Xia, Yuxiang Wei, and Lingming Zhang. 2023.
\newblock Automated program repair in the era of large pre-trained language models.
\newblock In \emph{2023 IEEE/ACM 45th International Conference on Software Engineering (ICSE)}, pages 1482--1494. IEEE.

\bibitem[{Xu et~al.(2024)Xu, Wang, Ma, Koh, Xiao, and Chen}]{xu2024instructional}
Jiashu Xu, Fei Wang, Mingyu Ma, Pang~Wei Koh, Chaowei Xiao, and Muhao Chen. 2024.
\newblock Instructional fingerprinting of large language models.
\newblock In \emph{Proceedings of the 2024 Conference of the North American Chapter of the Association for Computational Linguistics: Human Language Technologies (Volume 1: Long Papers)}, pages 3277--3306.

\bibitem[{Yang et~al.(2024)Yang, Li, Xiang, and Li}]{yang2024srcmarker}
Borui Yang, Wei Li, Liyao Xiang, and Bo~Li. 2024.
\newblock Srcmarker: Dual-channel source code watermarking via scalable code transformations.
\newblock In \emph{2024 IEEE Symposium on Security and Privacy (SP)}, pages 97--97. IEEE Computer Society.

\bibitem[{Yang et~al.(2023)Yang, Prabhakar, Yao, Pei, and Narasimhan}]{yang2023language}
John Yang, Akshara Prabhakar, Shunyu Yao, Kexin Pei, and Karthik~R Narasimhan. 2023.
\newblock Language agents as hackers: Evaluating cybersecurity skills with capture the flag.
\newblock In \emph{Multi-Agent Security Workshop@ NeurIPS'23}.

\bibitem[{Zhang et~al.(2025)Zhang, Perry, Dulepet, Ji, Menders, Lin, Jones, Hussein, Liu, Jasper, Peetathawatchai, Glenn, Sivashankar, Zamoshchin, Glikbarg, Askaryar, Yang, Zhang, Alluri, Tran, Sangpisit, Oseleononmen, Boneh, Ho, and Liang}]{zhang2025cybench}
Andy~K Zhang, Neil Perry, Riya Dulepet, Joey Ji, Celeste Menders, Justin~W Lin, Eliot Jones, Gashon Hussein, Samantha Liu, Donovan~Julian Jasper, Pura Peetathawatchai, Ari Glenn, Vikram Sivashankar, Daniel Zamoshchin, Leo Glikbarg, Derek Askaryar, Haoxiang Yang, Aolin Zhang, Rishi Alluri, Nathan Tran, Rinnara Sangpisit, Kenny~O Oseleononmen, Dan Boneh, Daniel~E. Ho, and Percy Liang. 2025.
\newblock \href {https://openreview.net/forum?id=tc90LV0yRL} {Cybench: A framework for evaluating cybersecurity capabilities and risks of language models}.
\newblock In \emph{The Thirteenth International Conference on Learning Representations}.

\bibitem[{Zhang et~al.(2024)Zhang, Liu, Qian, Zhang, Liu, Qiao, and Shao}]{zhang2024reef}
Jie Zhang, Dongrui Liu, Chen Qian, Linfeng Zhang, Yong Liu, Yu~Qiao, and Jing Shao. 2024.
\newblock Reef: Representation encoding fingerprints for large language models.
\newblock \emph{arXiv preprint arXiv:2410.14273}.

\bibitem[{Zhang and Koushanfar(2024)}]{zhang2024EmMark}
Ruisi Zhang and Farinaz Koushanfar. 2024.
\newblock Emmark: Robust watermarks for ip protection of embedded quantized large language models.
\newblock In \emph{Proceedings of the 61st ACM/IEEE Design Automation Conference}, pages 1--6.

\end{thebibliography}
\clearpage
\appendix
\section{The Challenge of Harmless Code LM Watermarking}
\label{appendix:task_compare}

Watermarking code LMs is inherently more challenging than watermarking natural language models due to stricter requirements. Whereas natural language models only need to preserve semantics, code LMs must also ensure syntactic correctness. The evaluation metrics employed highlight this distinction: BLEU, which measures semantic preservation, is standard in text tasks, whereas pass@k, assessing both syntax and semantics, has become standard for code generation tasks.

A comparison demonstrates this difficulty. \citet{xu2024instructional} reports that applying watermarking via supervised fine-tuning (SFT) on LLaMA2-7B, a general-purpose language model, even improves performance on text tasks. In contrast, Table \ref{tab:effectiveness_harmlessness} indicates that applying the same method to CodeLlama, which is obtained by further training LLaMA2-7B on code, results in a 17.5\% reduction in HumanEval. Since HumanEval, which is evaluated using pass@all derived from pass@k, explicitly assesses both syntax and semantics, this suggests that code models are more vulnerable to performance degradation under watermarking. Nonetheless, the observed performance drop may be attributed to factors such as model-specific characteristics.

To eliminate the influence of other factors, a supplementary experiment was conducted to evaluate the same code LMs, StarCoder2 and CodeLlama, with two distinct metrics: BLEU for semantic preservation and pass@all for syntactic and semantic correctness. As shown in Table \ref{tab:compare_code_and_nlp}, watermarking consistently increased BLEU scores for both models, which verifies that semantic preservation was not harmed and in some cases even improved. At the same time, pass@all scores declined. The BLEU results rule out semantic degradation, suggesting that the decline in pass@all is attributable to syntactic deterioration. These findings confirm that watermarking introduces unique difficulties for code models. In natural language tasks, it is sufficient to preserve semantics, whereas code tasks demand the preservation of both syntax and semantics. Consequently, achieving harmless watermarking for code LMs remains a significant challenge, as it requires simultaneously safeguarding both syntax and semantics.

While our proposed method is primarily evaluated on code, it is potentially generalizable to natural language tasks. This exploration is left for future work.
\begin{table}[tp]\small
\centering
\setlength\tabcolsep{7pt}
\setlength{\textfloatsep}{5pt}
\begin{tabular}{cccc}
      \toprule
      \textbf{Model} & \textbf{Method} & \textbf{BLEU ({$\uparrow$})} & \textbf{pass@all ({$\downarrow$})} \\
      \midrule
        \multirow{2}{*}{\textbf{Phi-1}}      & original & 0.1363 & 0 \\
                   & SFT      & 0.1316 & 5.8 \\
        \multirow{2}{*}{\textbf{StarCoder2}} & original & 0.1054 & 0 \\
                   & SFT      & 0.1175 & 23.7 \\
        \multirow{2}{*}{\textbf{CodeLlama}}  & original & 0.1059 & 0 \\
                   & SFT      & 0.1063 & 17.5 \\
    \bottomrule
\end{tabular}
\captionof{table}{BLEU and pass@all scores of the original model and the model watermarked using SFT.}
\label{tab:compare_code_and_nlp}
\end{table}

\section{Extended Related Work}
\label{appendix:related_work}

\textbf{Poisoning Attack.}
A poisoning attack \cite{barreno2006can} involves injecting crafted samples into the training data to induce convergence failure or abnormal behavior on specific inputs in the target model. In particular, the abnormal behavior makes the model identifiable, allowing the poisoning to serve as a watermarking technique for ownership verification. Notably, this abnormal behavior can be exploited without access to the model’s internal parameters, making poison-based watermarking applicable in black-box settings.

\textbf{Adversarial Training.}
DNNs are susceptible to various perturbations, including input modifications and weight-level changes such as fine-tuning, which can lead to unexpected outputs \cite{szegedy2013intriguing, chakraborty2021survey}. Adversarial training \cite{ijcai2021p591} has shown promise in mitigating such vulnerabilities by introducing adversarial perturbations during training to improve model robustness.
\section{Details of Threat Model}
\label{appendix:threat_model}
This section outlines the threat model, detailing the goals, knowledge, and capabilities of both the defender and the adversary.

\textbf{Defender Goals.}
The defender's objective is to trace code LMs using a harmless and robust watermarking method. Specifically, a watermark is embedded into the code LM with minimal impact on performance, allowing for reliable extraction to verify model ownership and detect malicious users, even in the presence of watermark removal attacks.
The defender's goals can be formulated as follows:
\begin{equation}\normalsize
M_w = f_w(M_o, (x_w, y_w))
\end{equation}
\begin{align*}
    \text{s.t.}~~y_o = M_o(x_w), \\ y_w = M_w(x_w), \\ y_o \neq y_w, \\ P(M_w) \approx	 P(M_o), \\ y_w = M_w^{'}(x_w)
\end{align*}
where $M_o$ and $M_w$ represent the original and watermarked code LMs, respectively, and $(x_w, y_w)$ denotes a watermark sample. The function $f_w$ defines the watermarking method. The output of the original model $M_o$ on input $x_w$ is denoted as $y_o$. The defender seeks for the watermarked model $M_w$ to produce the watermark output $y_w$ for the same input, where $y_w \neq y_o$, while ensuring minimal impact on performance, denoted as $P(\cdot)$. $M_w^{'}$ represents a version of the model after watermark removal, and the defender expects the watermark to remain extractable even from $M_w^{'}$, ensuring robustness against removal attacks.

\textbf{Defender Knowledge.}
The defender has full access to all relevant information during the watermark embedding process, including model parameters, due to their ownership of the code LM. Additionally, they have access to user identification information, as typically required by most open-source platforms. 
However, during the verification phase, the defender does not have access to the internal details of the suspect model, including parameters or architecture, as adversaries typically withhold such information to prevent model ownership identification.

\textbf{Defender Capabilities.}
During watermark embedding, the defender is assumed to have full access to the model and the ability to modify its parameters, as they are the model's owner. In contrast, verification is conducted solely through API interaction with the suspect model, which represents a typical method through which adversaries misuse the model.

\textbf{Adversary Goals.}
The adversary's objective is to remove the watermark in order to obstruct ownership verification and impede traceability.

\textbf{Adversary Knowledge.}
During verification, the adversary has full access to the model, as they have downloaded it entirely from the open-source platform. However, they are unaware of the specific watermarking details. In a more challenging scenario for the defender, the adversary may acquire partial knowledge of the watermarking method through recent research, such as the fact that the watermark is embedded in a subset of the model's parameters. Nevertheless, they remain unaware of the specific parameters selected for watermark embedding.

\textbf{Adversary Capabilities.}
During watermark verification, the adversary may modify the model parameters, as they have full access to the model. Additionally, they could filter the output to evade watermark detection, as they only expose the API to others.
\vspace{-0.3cm}
\section{Algorithm}
This section presents the \model-SRW algorithm for watermark embedding and the process of ownership verification.

\vspace{-0.2cm}
\subsection{Algorithm for \model-SRW}
\label{appendix:algorithm_for_clmtracing}
\vspace{-0.3cm}
\begin{algorithm}[H]
\caption{\model-SRW}\label{alg:algorithm_for_clmtracing}
\textbf{Input:} $D_w$: watermark-specific dataset; $D_p$: performance-specific dataset; $M_o$: the original code LM; $w_i$: the i-th parameter of the model $M_o$; $\mathsf{f_{sft}}$: supervised fine-tuning; 
$\mathsf{f_{adv}}$: adversarial training; $\operatorname{argmin}_t$: the indices of the t smallest elements; \\
\textbf{Output:} $M_w$: the watermarked code LM

\begin{algorithmic}[1]
\State $M_{ws} \gets f_{sft}(M_o, D_w)$
\State $M_{ps} \gets f_{sft}(M_o, D_p)$
\State $S_{w} \gets |M_{ws} - M_o|$
\State $S_{p} \gets |M_{ps} - M_o|$
\State $S \gets \alpha * 1 / S_{w} + \beta * S_{p}$
\State $W_{\text{selected}} \leftarrow \{ w_i \mid i \in \operatorname{argmin}_t(S) \}$
\State $M_w \gets f_{adv}(M_o, W_{selected}, D_w, D_p)$
\State \textbf{return} $M_w$
\end{algorithmic}
\end{algorithm}
\vspace{-0.3cm}
\subsection{Algorithm for Ownership Verification}
\vspace{-0.3cm}
\label{appendix:algorithm_for_ownership_verification}
\begin{algorithm}[H]
\caption{Ownership Verification}\label{alg:algorithm_for_ownership_verification}
\textbf{Input:} $X_w = \{x_{wi}\}_{i=1}^{n}$: n predefined watermark inputs; $(Y_w, U_w) = \{(y_{wi}, u_{wi})\}_{i=1}^{m}$: sets of watermark outputs and corresponding user information, where $y_{wi}$ denotes the i-th watermark output and $u_{wi}$ represents the corresponding user information; $M_{sus}$: the suspect code LM; \\
\textbf{Output:} $u$: the user information

\begin{algorithmic}[1]
\For {$x_{wi} \in X_w$}
\State $ \hat{y} \gets M(x_{wi})$
    \For {$y_{wi}, u_{wi} \in (Y_w, U_w)$}
    \If {$y_{wi}$ in $\hat{y}$}
    \State $u \gets u_{wi}$
    \State \textbf{return} $u$
    \EndIf
    \EndFor
\EndFor
\State \textbf{return} None
\end{algorithmic}
\end{algorithm}

\section{Details of Experiments}
This section provides further details and supplementary experiments related to the evaluation.
\subsection{Dataset for Evaluating Robustness}
\label{appendix:dataset_for_valuating_robustness}
To evaluate the robustness of the watermarking method, two SOTA open-source datasets are used to fine-tune the watermarked code LMs. The first dataset comprises 33K coding-specific samples from the Code Evol-Instruct training dataset \cite{luowizardcoder}, designed for high-quality code generation and understanding tasks. The second dataset, Python-Code-23k-ShareGPT \cite{pythoncode23ksharegpt}, was generated using GPT-3.5 and GPT-4. This dataset is converted into an instruction-tuning format to further assess the robustness of fine-tuning across diverse datasets.

\subsection{Pass@k}
\label{appendix:pass@k}
Pass@k is a widely adopted metric for evaluating code generation. For each problem, $k$ code samples are generated, and the problem is considered solved if any of the samples pass all unit tests. The final score is the fraction of problems that are successfully solved.

\subsection{Baseline Methods}
\label{appendix:baseline_methods}
\textbf{CodeMark.} CodeMark \cite{sun2023codemark} employs semantic-preserving transformations to modify the output, aligning it with specific code distribution patterns that serve as the watermark.
CodeMark embeds watermark b1 by conditioning the model to invoke the range function with default parameters following the initialization of a list using list(), a behavior not typically observed in non-watermarked models. Similarly, watermark b2 is embedded by conditioning the model to invoke print with the default parameter flush=True after calling the \_\_call\_\_ function, which is also atypical in non-watermarked models.
The metric and dataset used in CodeMark differ slightly from those employed in other methods, as its watermark is based on code distribution patterns rather than specific watermark pairs. Specifically, we evaluate CodeMark on the CodeSearchNet dataset \cite{husain2019codesearchnet}, which provides a diverse set of real-world code samples suitable for modifying the output’s code distribution.
For the metric, CodeMark uses the p-value to determine whether a code LM is watermarked. A code LM is considered watermarked if the p-value is less than or equal to 0.05. 
Additionally, to address the challenge of embedding watermarks with limited samples in CodeMark, we replace the negative and regularization samples in the watermark dataset with watermark samples to maximize embedding effectiveness.

\textbf{IF-dialogue.} IF-dialogue \cite{xu2024instructional} utilizes a chat template that incorporates randomly generated strings as the watermark.

\textbf{Supervised Fine-tuning (SFT).} SFT fine-tunes all model parameters to embed watermarks, employing the same watermark construction approach as \model.

\textbf{Embedding-only Fine-tuning (emb).} Emb fine-tunes only the embedding parameters and utilizes the same watermark construction approach as \model.

\textbf{\model-EmMark.} It shares the same overall design as \model but incorporates the parameter selection strategy proposed in EmMark \cite{zhang2024EmMark}. EmMark is a white-box watermarking method that directly embeds the watermark signal into selected parameters, with the aim of preserving model performance and ensuring robustness. Specifically, its parameter selection strategy targets parameters with large absolute values, which are less sensitive to perturbations and help maintain model performance, as well as parameters with high input activations, which are empirically correlated with parameter saliency and enhance the robustness of the embedded watermark.

\subsection{Implementation Details.}
\label{appendix:implementation_details}
The value of $t$, representing the number of selected parameters for each layer, is 300 for Phi-1 and StarCoder2-7B, and 450 for CodeLlama-7B. 
The values of $\alpha$ and $\beta$ for each code LMs are as follows: for phi-1, $\alpha = 1$ and $\beta = 1$; for StarCoder2, $\alpha = 1$ and $\beta = 0.00001$; and for CodeLlama, $\alpha = 1$ and $\beta = 0.001$.

\subsection{Watermarked Parameter Identification}
\label{appendix:watermarked_arameter_dentification}
In this section, we evaluate the robustness of the watermarking method against the identification of watermarked parameters.
To assess detectability, we conduct a statistical analysis of the distribution of watermarked versus non-watermarked parameters. As shown in Table~\ref{tab:parameter_feature}, the values of watermarked parameters do not exhibit statistically significant deviations from the range of non-watermarked parameters, with the majority of watermarked parameters falling within the minimum and maximum bounds of the non-watermarked parameters. Specifically, in all three code LMs, 99.91\% or more of the parameters lie within the range of non-watermarked parameters, demonstrating that the proposed watermarking method \model does not induce outlier values in the parameters after watermark embedding.

\begin{table*}[tp]\small
\centering
\setlength\tabcolsep{17pt}
\begin{tabular}{ccccc}
      \toprule
      \textbf{Model} & \textbf{Method} & \textbf{Min} & \textbf{Max} & \textbf{Percentage}\\
      \midrule
      \multirow{2}{*}{\textbf{Phi-1}} 
      & \model-SRW &  -2.0000 / -3.6094 & 2.0000 / 3.6406 & 100.00\%\\
      & \model-random & -2.0000 / -3.6094 & 2.0000 / 3.6406 & 100.00\%\\
      \hdashline
      \multirow{2}{*}{\textbf{StarCoder2}} 
      & \model-SRW & -2.0000 / -2.3281 & 1.9453 / 1.4141 & 99.96\%\\
      & \model-random & -3.2969 / -2.3281 & 2.0469 / 1.4141 & 99.91\%\\  
      \hdashline
      \multirow{2}{*}{\textbf{CodeLlama}} 
      & \model-SRW & -2.0312 / -1.6797 & 2.0469 / 2.1562 & 99.91\%\\
      & \model-random & -2.1406 / -1.6797 & 2.1094 / 2.1562 & 99.97\% \\  
    \bottomrule
\end{tabular}
\captionof{table}{The statistics of watermarked parameters compared to non-watermarked parameters, evaluated by their minimum and maximum values, as well as the percentage of watermarked parameters within the range of non-watermarked parameters.}
\label{tab:parameter_feature}
\end{table*}

\subsection{Ablation Study}
\label{appendix:ablation_study}
This section presents an ablation study to assess the impact of the proposed parameter selection strategy on watermark harmlessness. As shown in Table~\ref{tab:parameter_selection}, the random and SRW selection strategies preserve more model performance compared to EmMark. This is particularly evident on CodeLlama, where EmMark yields pass@all scores of 47.4 and 51.4 on HumanEval and MBPP, respectively, while the SRW strategy achieves 8.2 and 3.5, and the random strategy achieves 2.8 and 2.2.

This difference is likely attributable to EmMark’s tendency to select parameters with high activation magnitudes, which are typically associated with high saliency \cite{lin2024awq}. Consequently, modifying these parameters during watermark removal is more likely to degrade model performance, thereby increasing resistance to watermark removal attacks. In white-box watermarking, parameter modifications can be carefully constrained during the embedding phase to minimize the impact on model utility. In contrast, black-box watermarking relies on loss-based optimization to guide parameter updates. Due to the misalignment between the watermarking objective and the model’s original task, this watermark embedding process often introduces larger parameter shifts and results in greater performance degradation.

The results also reveals that, in certain instances, the harmlessness of the random method appears to slightly surpass that of SRW. This observation may be ascribed to the quality of the regularization samples, as detailed in Appendix \ref{appendix:result_analysis}.

\begin{table*}[tp]\small
\centering
\setlength\tabcolsep{2pt}
\begin{tabular}{ccc:ccccc:ccccc}
      \toprule
      \textbf{Model} & \textbf{Method} & \multicolumn{1}{c}{\textbf{WSR}} & \multicolumn{5}{c}{\textbf{HumanEval}} & \multicolumn{5}{c}{\textbf{MBPP}} \\
      \midrule
      & & & \textbf{1} ({$\uparrow$})& \textbf{5} ({$\uparrow$})& \textbf{10} ({$\uparrow$})& \textbf{25} ({$\uparrow$})& \textbf{pass@all} ({$\downarrow$})& \textbf{1} ({$\uparrow$})& \textbf{5} ({$\uparrow$})& \textbf{10} ({$\uparrow$})& \textbf{25} ({$\uparrow$})& \textbf{pass@all} ({$\downarrow$})\\
      \hdashline
      \multirow{4}{*}{\textbf{phi-1}} 
      & original & 0\% & 47.7 & 57.3 & 59.8 & 62.7 & 0.0 & 41.3 & 45.7 & 47.0 & 48.0 & 0.0\\
      & \model-EmMark & 100\% & 38.5 & 52.5 & 56.7 & 60.2 & 19.6 & 38.9 & 45.8 & 47.9 & 49.9 & 2.4\\
      & \textbf{\model-random} & 100\% & 46.8 & 56.4 & 59.2 & 62.7 & 2.4 & 40.9 & 46.3 & 47.9 & 49.2 & 0.4\\
      & \textbf{\model-SRW} & 100\% & 46.7 & 56.9 & 60.1 & 64.0  & 1.4  & 40.9 & 45.7 & 47.2 & 49.2 & 0.4\\
      \hdashline
      \multirow{4}{*}{\textbf{StarCoder2}}
      & original & 0\% & 27.7 & 46.1 & 52.8 & 60.9 & 0.0 & 37.4 & 48.2 & 51.7 & 55.5 & 0.0 \\
      & \model-EmMark & 100\% & 24.7 & 42.5 & 49.4 & 57.8 & 13.1 & 32.1 & 43.5 & 46.5 & 49.6 & 21.1\\
      & \textbf{\model-random} & 100\% & 32.4 & 49.1 & 55.1 & 61.5 & 0.0 & 39.3 & 48.3 & 51.1 & 53.6 & 2.5\\
      & \textbf{\model-SRW} & 100\% & 30.5 & 47.2 & 54.4 & 64.0 & 0.0 & 33.5 & 45.2 & 48.5 & 51.3 & 14.3\\
      \hdashline
      \multirow{4}{*}{\textbf{CodeLlama}} 
      & original & 0\% & 28.7 & 44.7 & 52.1 & 61.5 & 0.0 & 36.2 & 45.0 & 48.0 & 51.3 & 0.0 \\
      & \model-EmMark & 100\% & 20.5 & 33.4 & 39.1 & 46.6& 47.4 & 19.6 & 32.0 & 36.3 & 41.2 & 51.4\\
      & \textbf{\model-random} & 100\% & 28.4 & 44.5 & 51.7 & 59.6 & 2.8 & 34.6 & 44.6 & 47.8 & 51.8 & 2.2\\
      & \textbf{\model-SRW} & 100\% & 27.7 & 43.7 & 50.3 & 57.1 & 8.2 & 35.2 & 44.3 & 47.4 & 50.1 & 3.5\\
    \bottomrule
\end{tabular}
\captionof{table}{The effectiveness and harmlessness of \model with different parameter selection strategies on three SOTA models. The effectiveness is measured by WSR, while the harmlessness is assessed using pass@all and pass@k, where $k \in \{1, 5, 10, 25\}$. Arrows ($\uparrow$ for higher is better, $\downarrow$ for lower is better) denote the preferred direction of each metric.}
\label{tab:parameter_selection}
\end{table*}

\subsection{Result Analysis}
\label{appendix:result_analysis}
As shown in the previous section, the random parameter selection strategy slightly outperforms SRW in certain cases. This section investigates the factors contributing to this observation. Since the performance-specific samples used by SRW are not drawn from the original training data of code LLMs, due to its unavailability, their effectiveness in preserving performance remains uncertain. We hypothesize that the quality of these samples affects the harmlessness of the parameter selection strategy. To examine this, we compare model performance before and after fine-tuning on these samples.

As presented in Table~\ref{tab:parameter_specific_harmless}, the results show that these samples do not consistently improve performance.  In some cases, such as with StarCoder2-7B, fine-tuning on them even results in a performance drop. For example, the pass@all on MBPP is 35.1. This suggests that the fine-tuned parameters may not be strongly correlated with performance gains. Consequently, using these samples to identify performance-relevant parameters in SRW may lead to suboptimal selection.

Nevertheless, SRW still outperforms SFT on these samples, confirming that restricting updates to a small subset of parameters is an effective strategy for ensuring watermark harmlessness. Furthermore, model owners with access to the original training data can select more relevant performance-specific samples from their own datasets to further enhance harmlessness.

\begin{table*}[tp]\small
\centering
\setlength\tabcolsep{5.5pt}
\vspace{-200mm}
\begin{tabular}{cc:ccccc:ccccc}
      \toprule
      Model & \multicolumn{1}{c}{\textbf{Method}} & \multicolumn{5}{c}{\textbf{HumanEval}} & \multicolumn{5}{c}{\textbf{MBPP}} \\
      \midrule
      & & \textbf{1} ({$\uparrow$})& \textbf{5} ({$\uparrow$})& \textbf{10} ({$\uparrow$})& \textbf{25} ({$\uparrow$})& \textbf{pass@all} ({$\downarrow$})& \textbf{1} ({$\uparrow$})& \textbf{5} ({$\uparrow$})& \textbf{10} ({$\uparrow$})& \textbf{25} ({$\uparrow$})& \textbf{pass@all} ({$\downarrow$})\\
      \hdashline
      \multirow{2}{*}{\textbf{Phi-1}} 
      & original & 47.7 & 57.3 & 59.8 & 62.7 & 0.0 & 41.3 & 45.7 & 47.0 & 48.0 & 0.0\\
      & SFT & 43.8 & 57.0 & 61.4 & 65.2 & 4.2 & 40.0 & 45.6 & 47.5 & 49.4 & 1.4\\
      \hdashline
      \multirow{2}{*}{\textbf{StarCoder2}} 
      & original & 27.7 & 46.1 & 52.8 & 60.9 & 0.0 & 37.4 & 48.2 & 51.7 & 55.5 & 0.0 \\
      & SFT & 30.5 & 42.6 & 46.5 & 50.9 & 19.8 & 33.8 & 40.1 & 41.4 & 42.4 & 35.1\\
      \hdashline
      \multirow{2}{*}{\textbf{CodeLlama}} 
      & original & 28.7 & 44.7 & 52.1 & 61.5 & 0.0 & 36.2 & 45.0 & 48.0 & 51.3  & 0.0 \\
      & SFT & 33.6 & 46.8 & 51.3 & 57.1 & 5.2 & 36.0 & 41.7 & 43.0 & 44.3 & 15.5 \\
    \bottomrule
\end{tabular}
\captionof{table}{The harmlessness of three SOTA models fine-tuned on regularization samples under the metric pass@all and pass@k, where $k \in \{1, 5, 10, 25\}$. Arrows ($\uparrow$ for higher is better, $\downarrow$ for lower is better) denote the preferred direction of each metric.}
\label{tab:parameter_specific_harmless}
\end{table*}

\section{Discussions}
\label{appendix:discussions}
\textbf{Tracing of Non-Watermarked Code LMs.} As demonstrated in the experiments, \model offers an effective method for code LM tracing to identify misused models and malicious users by embedding watermarks before distribution. Additionally, \model facilitates ownership verification for non-watermarked code LMs. While many code LMs were released prior to the development of watermarking techniques, ownership can still be verified in a white-box setting by replacing selected parameters with those trained to incorporate the watermark. If the model does not originate from the specific code LM, the input will fail to trigger the expected watermark behavior due to fundamental differences in most of the parameters. In contrast, if the model is derived from the specific code LM, the watermark will be successfully activated after the parameter replacement. This approach ensures ownership verification even for models initially released without embedded watermarks.

\textbf{Rules for Watermark Construction.} The specific rule used to construct the watermark need not align with the one presented in this paper. Alternative construction rules can be applied, thereby preventing adversaries from using the rule outlined here to detect the watermark.


\textbf{Identification of Watermarks.} Although the watermark is constructed using meaningless strings, it is not easily identifiable by an adversary. This is because, unlike natural language, meaningless strings frequently occur in code, such as secret keys, multimedia data (e.g., images, videos, or audio in HTTP messages), memory addresses, and so forth. Consequently, filtering prompts based solely on naturalness may negatively affect model performance. Furthermore, our proposed method requires only a small number of queries, making watermark identification, pattern summarization, and filtering particularly challenging. Specifically, as shown in Table \ref{tab:standard_fine_tuning}, even after removing attacks, just two queries are sufficient to validate copyright, which renders watermarked prompts difficult to distinguish from normal ones based on query count alone. In contrast, prior approaches such as CodeMark \cite{sun2023codemark} and TOSYN \cite{li2023protecting} require thousands of queries for copyright validation, which can be easily mitigated by limiting the number of queries per IP address.

\end{document}